\documentclass{article} % For LaTeX2e
\usepackage[preprint]{colm2026_conference}

\usepackage{microtype}
\usepackage{hyperref}
\usepackage{url}
\usepackage{booktabs}
\usepackage{graphicx} 
\usepackage{threeparttable} % in preamble
\usepackage{tcolorbox}
\usepackage{caption}
\tcbuselibrary{breakable}
\usepackage{listings}
\usepackage{lineno}

\definecolor{darkblue}{rgb}{0, 0, 0.5}
\hypersetup{colorlinks=true, citecolor=darkblue, linkcolor=darkblue, urlcolor=darkblue}

\title{Revisiting Framing Codebooks with AI: Employing Large Language Models as Analytical Collaborators in Deductive Content Analysis}

\author{Diego Gómez-Zará \\
Department of Computer Science and Engineering, University of Notre Dame, USA \\
\texttt{[dgomezara@nd.edu]} \\
\And
Hernán Valdivieso \\
Department of Computer Science \\
Pontificia Universidad Católica de Chile, Chile \\
\texttt{[hfvaldivieso@uc.cl]} \\
\And
Jorge Pérez \\
Cero.ai \\
Santiago, Chile \\
\texttt{[jorge.perez.rojas@gmail.com]} \\
\And
Denis Parra \\
Department of Computer Science, Pontificia Universidad Católica de Chile, Chile \\
Millennium Institute for Foundational Research on Data, Chile \\
\texttt{[dparras@uc.cl]} \\
\And
Sebastián Valenzuela \\
Facultad de Comunicaciones, Pontificia Universidad Católica de Chile, Chile \\
Millennium Institute for Foundational Research on Data, Chile \\
Millennium Nucleus on Digital Inequalities and Opportunities, Chile \\
\texttt{[savalenz@uc.cl]}
}

\begin{document}

\ifcolmsubmission
\linenumbers
\fi

\maketitle

\begin{abstract}
Codebooks are central to framing research, providing theoretically grounded criteria for analyzing news content. While traditionally codebooks are built from theoretical frameworks and researchers' knowledge, applying these codebooks to large news corpora often exposes ambiguities, borderline cases, and underspecified rules that are difficult to resolve through theory alone. Moreover, news corpora evolve over time and differ across cultures, necessitating that researchers revisit the theoretical frameworks underlying these codebooks. In this article, we propose a workflow that uses Large Language Models (LLMs) to augment the creation and refinement of framing codebooks by combining theoretical frameworks with data-driven exploration. Rather than treating LLMs as automated classifiers, this approach positions them as analytic collaborators that help externalize decision rules, surface latent dimensions, and support iterative revisions of codebooks through dialogues between researchers and their data. We illustrate this workflow using a dataset of Latin American news coverage, demonstrating how the application of LLMs' capabilities has led to the surfacing of latent patterns, the generation of frame distinctions, and the adaptation of frameworks to new contexts. This method provides an LLM-assisted strategy that supports methodology creativity while preserving researchers' interpretative authority. 
\end{abstract}

\section{Introduction}
Digital journalism scholars have extensively discussed the implications of artificial intelligence (AI) and, more recently, large language models (LLMs) on journalists' core practices \citep{Arguedas2023-qy,Broussard2019-kd,Simon2024-gz,Pilny2024-jv}. These technologies can automate journalists' writing tasks \citep{Agrawal2023-wl}, generate diverse angles for covering news events \citep{Petridis2023-wf}, enhance fact-checking workflows \citep{Choi2024-mk}, and support tasks such as summarization, newsgathering, and brainstorming \citep{Nishal2023-mx,Nishal2024-bn}. 

Scholars across disciplines have also explored how LLMs can enhance research practices, including generating ideas \citep{Baek2024-ve}, literature reviews \citep{Dennstadt2024-pc,Zimmermann2024-et}, hypotheses \citep{Hope2023-ja,Zhou2024-re}, and assisting scholarly writing \citep{Gero2022-qc,Koller2024-cz}. Despite these advances, the methodological potential for LLMs in journalism research has received less attention \citep{Sarisakaloglu2025-ms}. This gap is consequential given journalism studies' reliance on intensive iterative judgment and interpretation. 

In this advancing methods article, we propose a workflow that positions LLMs not as automated classifiers, but as interactive analytical collaborators that support journalism researchers in the creation, interrogation, and refinement of deductive codebooks used for framing analysis. Although codebooks are typically grounded in established theoretical frameworks and researchers' domain knowledge, applying them to large, highly contextual, time-dependent, and heterogeneous news corpora often reveals ambiguities, borderline cases, and underspecified evaluation criteria that are difficult to anticipate in advance. Rather than treating these challenges as errors, we view them as opportunities for methodological and theoretical refinement. Unlike prior computational approaches that rely on model training or fixed feature extraction, this method enables researchers to engage directly with LLMs, using theoretical definitions, contextual examples, and decision rules in response to empirical patterns identified in the data. 

We illustrate how this LLM-based coding approach draws on a framing study on news articles published in a Latin American media outlet \citep{Valenzuela2023-dy}. The workflow enables researchers to iteratively revise codebooks in response to empirical patterns surfaced through interaction with LLMs, while retaining interpretative authority. In doing so, it supports methodological creativity in theory-driven content analysis by shifting codebook development from a one-time, upfront task to an iterative, dialogic process in which theoretical assumptions are surfaced, tested, and revised.

\section{Background}
Journalism research has traditionally relied on trained annotators for manual framing analysis \citep{Krippendorff2018-dn,Neuendorf2017-tz}. As news volume has increased substantially over the past decades, researchers have turned to computational techniques for larger-scale data analysis \citep{Burscher2015-vi,Garcia-Marin2018-tx,Baden2022-hg,Boumans2016-nu}. While supervised machine learning methods have supported deductive coding strategies by predicting frames from trained datasets \citep{Garcia-Marin2018-tx,De_Grove2020-ng,Burscher2014-jl}, unsupervised machine learning techniques (e.g., topic modeling) have supported inductive strategies by surfacing latent frames in the data \citep{Scharkow2013-fx,Walter2019-kf,Guo2016-gw}. Machine learning has expanded the methods used in journalism research. Vector embeddings enable more context-sensitive analysis of news content \citep{Kroon2022-ro,Kroon2021-sx}. Other approaches, such as the Analysis of Topic Model Networks (ANTMN), combine topic modeling with network analysis to interpret frames within large news corpora \citep{Walter2019-kf,Walter2020-qc}. Similarly, Guo et al. \citeyearpar{Guo2023-ep} integrate topic modeling with deep learning to study the co-occurrence of multiple frames. 
Despite these advances, many approaches require substantial technical expertise in coding, data preprocessing, and model tuning, limiting researchers' ability to directly inspect or revise analytic decisions \citep{Lazer2020-jq,Edelmann2020-yy}. More importantly, these approaches offer limited support for iterative, dialogic revision of deductive codebooks. Feedback from models is typically indirect and aggregate, making it difficult to systematically trace how specific cases expose ambiguities or strain theoretical assumptions embedded in coding schemes \citep{Baden2022-hg,Van_Atteveldt2021-dg,Grimmer2021-mf}  

LLMs introduce new opportunities for interacting with natural language representations of theories, criteria, and data. By predicting likely subsequent word sequences from a text prompt, they enable analysis, generation, and reasoning that approximate human interpretive practices \citep{Chang2024-as,Kasneci2023-ge}. LLMs can integrate both pre-trained knowledge and researchers' knowledge and datasets, and—unlike traditional machine learning methods—allow iterative feedback through dialogue to align model outputs with researchers' analytic goals \citep{Min2023-qb,Wang2023-av}. Consistent with this potential, Pilny et al. \citeyearpar{Pilny2024-jv} show that LLMs can produce classifications comparable to human coding, supporting their use in nuanced content analysis. Building on these capabilities, researchers have applied LLMs to both inductive and deductive coding across document, social media, and news content analysis \citep{Gilardi2023-ml,Xiao2023-nx,Fan2024-te}. Using LLMs, several studies have increasingly examined human-AI workflows that support interactive inductive coding, rule synthesis, and collaborative sensemaking \citep{Gebreegziabher2023-pu,Gao2024-hl,Chew2023-xt,Than2025-zn}.

However, much of this work has been oriented toward benchmarking performance or supporting general-purpose qualitative workflows, rather than addressing specific challenges journalism scholars face when applying and revising theory-driven deductive codebooks across large and heterogeneous news corpora. Questions about how LLMs can support corpus-level exploration, surface borderline cases, and facilitate refinement of evaluation criteria remain particularly salient \citep{Gebreegziabher2025-lv}. 

\section{A Step-by-Step Guide}
\begin{figure}
    \centering
    \includegraphics[width=\linewidth]{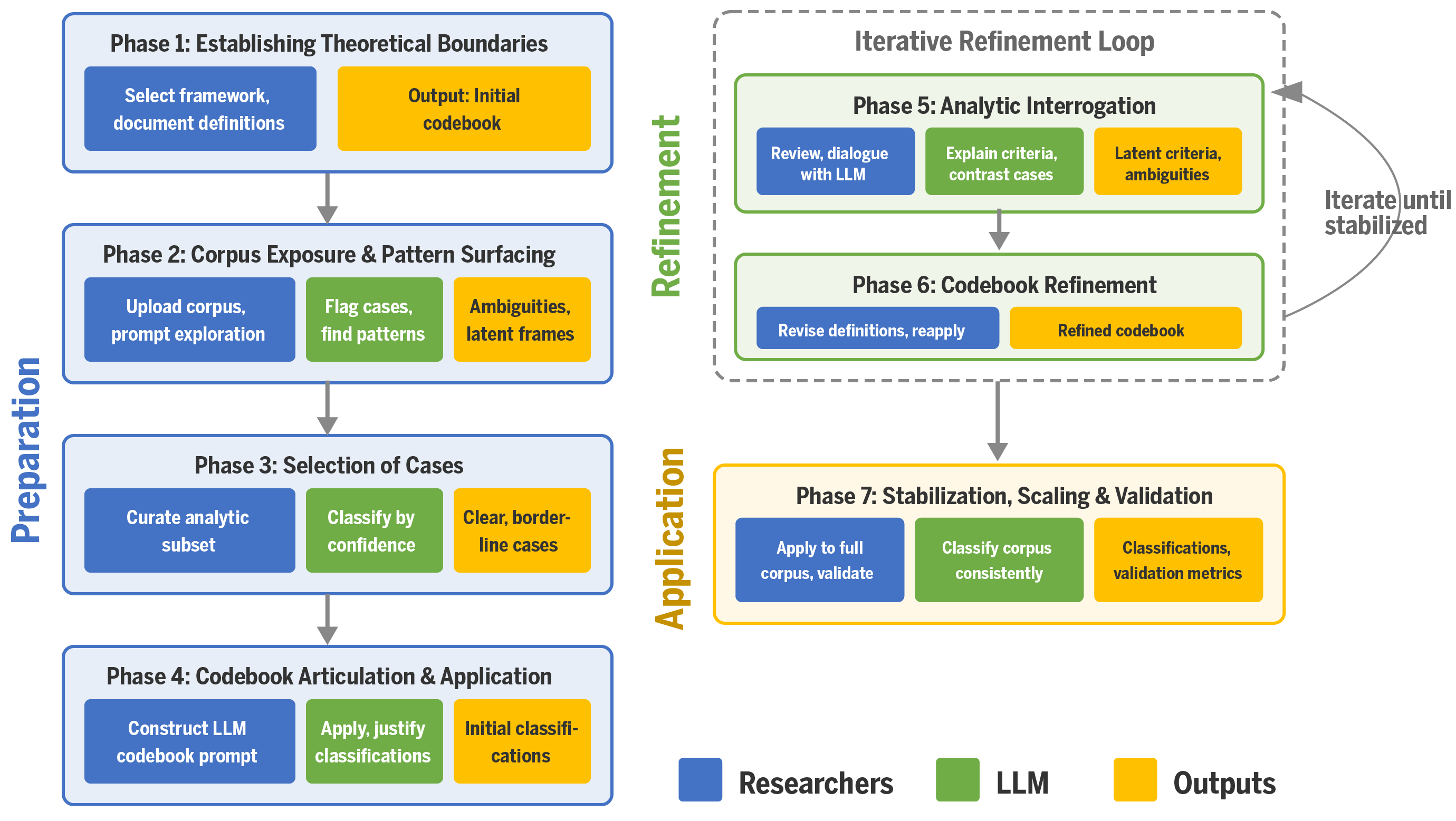}
    \caption{Overview of the LLM Codebook Prompt}
    \label{fig:overview}
\end{figure}

Building on this literature, we outline a step-by-step workflow for integrating LLMs into journalism research that supports methodological creativity in the development and application of theoretically grounded codebooks. The following section details each phase—from initial corpus exposure to codebook stabilization—and illustrates how researchers can interact with LLMs to surface ambiguities, examine borderline cases, and revise coding criteria in response to empirical tensions, while retaining interpretative authority (see Figure \ref{fig:overview}). Rather than treating the codebook as fixed once analysis begins, the workflow frames codebook development as an iterative, dialogic process in which theoretical assumptions are made explicit, contested, and refined through engagement with the data. This mirrors how deductive codebooks are typically developed in qualitative and content research, where human coders are involved, and disagreements and edge cases prompt clarification and revision. Here, this analytic practice is enacted through interaction with an LLM, which externalizes rationales, highlights alternative category boundaries, and supports systematic theoretical refinement on a large scale. 

\subsection{Phase 1. Establishing Theoretical Boundaries}
Researchers begin by selecting a theoretical framework and specifying the deductive frames for the analysis. These frames may be drawn directly from the original coding manuals, adapted from prior empirical studies, or developed to suit a new research context. Researchers assemble illustrative examples to clarify intended interpretations and explicitly document frame definitions, assumptions, and analytical goals, forming an initial codebook. 
For instance, researchers using the episodic and thematic framing framework \citep{Iyengar1994-pu} might compile canonical definitions, decision rules, and example passages that distinguish between episodic from thematic coverage. These elements are treated as initial commitments that will later be interrogated and refined through interaction with the corpus and the LLM.

\subsection{Phase 2. Broad Corpus Exposure and Pattern Surfacing}
Researchers interact with the LLM using the initial codebook, with exploratory rather than classificatory goals, examining how frames operate across the corpus and identifying frame classifications that are clear, ambiguous, or strained. Researchers expose the LLM to a substantial portion of the corpus, either the full dataset with specific features (e.g., headlines and leads) or a representative stratified subset (e.g., 20-50\% of articles). Using the initial frame definitions, researchers prompt the LLM to tentatively identify the prevalence of these frames in the corpus, uncover potential frames, and flag uncertain cases. By interacting with the LLM, researchers can request summaries of recurring rationales and identify clusters that merit closer examination.

For example, using Iyengar's thematic framing framework for political news coverage, researchers provide the LLM with a collection of headlines and leads from the corpus and ask it to assess whether coverage is episodic or thematic. During this initial scan, the LLM may repeatedly mark articles that describe individual events (e.g., a single protest, a court ruling, a policy announcement), but also include brief references to broader trends or historical context as ambiguous. In its explanations, the model might note that while these articles are anchored in specific events, they also gesture toward structural cases without fully developing them.

\subsection{Phase 3. Selection of Representative and Borderline Cases}
Building on the patterns surfaced in Phase 2, researchers ask the LLM to curate a subset of articles for close analysis, assembling cases that capture how the theoretical framework operates across clear, ambiguous, and difficult instances. These examples disambiguate constructs by presenting concrete cases and grounding the rationale for decisions. Cases may include contextual information retrieved from the dataset—such as the source, country, public figures, or time period—to better ground the LLM's interpretations. 

Following the illustrative example, the LLM identifies a clear episodic case, focusing on a single event with no broader contextualization; a borderline case centered on a policy decision with undeveloped references to long-term trends; and an unclear case, summarizing procedural developments with minimal narrative framing. These articles serve as anchor examples for subsequent phases, grounding codebook articulation in concrete instances rather than relying solely on abstract definitions.

\subsection{Phase 4. Initial Codebook Prompt and Application}
Researchers translate the previsional LLM codebook and curated cases into an initial prompt operationalizing the theoretical framework. The goal is to make theoretical assumptions explicit and test how frames are enacted across representative cases. Researchers construct an initial ``LLM codebook prompt'' that specifies frame definitions, analytic questions, and illustrative examples. As with training human annotators, the prompt requires precise instructions about how frames should be interpreted. When available, the suggested cases can be complemented with previously annotated data or established coding manuals. The prompt should request the LLM to explain its reasoning for each classification, allowing decisions to be inspected and revised later. The prompt commonly includes the following components (See Figure \ref{fig:meta-prompt} for a simplified example):
\begin{itemize}
    \item \textit{Role instruction (Optional).} It suits the LLM in an analytic perspective (e.g., ``You are a communication researcher expert in framing theory...'')
    \item \textit{Frame definitions}, which are explicitly referenced from the theoretical framework. Researchers could include the original citation, add the definitions, or adapt them based on their interpretation.
    \item \textit{Case Examples,} both positive and negative, to help the LLM narrow and guide the classification through ``few-shot'' prompting \citep{Brown2020-ds,Xiao2023-nx}.
    \item \textit{Framing questions}, instructing the model to assess the presence or absence of specific frames, and justifying its assessment.
    \item \textit{Article text}, clearly delimited within the prompt or provided as an attachment.
    \item \textit{Output instructions}. This instruction outlines the structure of responses to ensure consistency across cases.
\end{itemize}

\begin{figure}[!htb]
\centering
\begin{tcolorbox}[
    colback=gray!5,
    colframe=black,
    width=\linewidth,
    arc=2mm,
    boxrule=0.5pt,
    breakable
]
\scriptsize
\textbf{Role Instruction:} \\
You are an expert assistant for content analysis...

\vspace{4pt}
\textbf{Theoretical Framework / Codebook Explanation:} \\
You will apply a codebook of frames to a news article and code the article accordingly. The framework is \texttt{\{framework\_name\}} and its citation is \texttt{\{framework\_citation\}}.

\vspace{4pt}
\textbf{Frames and Definitions:} \\
For each frame, include: definition, include/exclude rules, and examples/counterexamples.

\vspace{4pt}
\textbf{Instructions:} \\
Read the article below. Based on the definitions, determine which frame(s) the article uses. Provide the code label and a brief justification, quoting or paraphrasing evidence from the article to support your decision. Do not inject information not described in the article.

\vspace{4pt}
\textbf{Article to Analyze:} \\
\textit{"Full text of the news article goes here"}

\vspace{4pt}
\textbf{Output Instructions:} \\
When you finish answering, generate a spreadsheet file with the following format...
\end{tcolorbox}
\caption{Prompt used for frame classification.}
\label{fig:meta-prompt}
\end{figure}

\subsection{Phase 5. Analytic Interrogation and Criteria Elicitation}
Using the initial codebook, researchers systematically interrogate the LLM's explanations to surface implicit decision rules, latent criteria, and classification inconsistencies. Rather than evaluating outputs solely for accuracy, the goal is to examine how the LLM interprets the codebook and where its reasoning reveals ambiguities or tensions within the theoretical framework. Researchers review justifications across the analytical subset, attending to recurring rationales, shifts in emphasis, and cases where similar features yield different classifications. Patterns, such as repeated references to specific cues, inconsistent weighting of contextual information, or expressions of uncertainty, can indicate underspecified coding rules or latent distinctions not explicit in the initial codebook. These summaries are treated as candidate articulations of implicit decision rules and latent criteria, which researchers could accept, revise, or reject. When available, manually annotated cases can be used to diagnostically compare the LLM's reasoning against researchers' expectations.

Continuing the episodic–thematic framing study, researchers examine the LLM's justification for a borderline case. The model explains its uncertainty by noting that contextual references are present but brief and lack causal development. When prompted to summarize its reasoning, it distinguishes between passing mentions and structural explanations, revealing a latent criterion not explicit in the original codebook that helps clarify the episodic–thematic boundary.

\subsection{Phase 6. Codebook Refinement and Theoretical Clarification}
Researchers revise the codebook in response to the criteria, ambiguities, and suggestions surfaced through the interrogations. Revisions may involve refining frame definitions, clarifying decision rules, adding or adjusting examples, or explicitly incorporating latent distinctions identified in the LLM's reasoning. Researchers update the prompt by rewording framing questions, expanding illustrative examples, or specifying how particular cues should be weighted. When explanations are overly brief, vague, strict, or inconsistent, researchers should also revise instructions to elicit more targeted reasoning. 

The revised codebook is reapplied to the analytical subset to determine whether ambiguities have been resolved. If manually annotated data are available, researchers compare classifications to identify where human and LLM evaluations diverge. Iteration continues until applying the revised codebook no longer yields new criteria, disagreements, or recurring ambiguities. By the end of this process, researchers arrive at a stabilized, high-fidelity codebook with clear definitions, broad coverage, explicit decision rules, and representative examples that remain grounded in the original theoretical framework.

Continuing the episodic–thematic framing study, researchers incorporate the distinction between passing mentions and structural explanations into the codebook. The thematic frame is revised to require sustained causal or structural contextualization, and an additional example is added to illustrate this criterion. When the revised prompt is applied to the analytic subset, previously ambiguous articles are classified more consistently, with the LLM's justifications explicitly referencing the revision.

\subsection{Phase 7. Stabilization, Scaling, and Validation}
The final phase marks the transition from the iterative refinement to a stabilized application. Once the codebook is no longer in need of updates, researchers apply the resulting prompt consistently to the entire corpus. Researchers can apply the finalized prompt to each article with human evaluators, or with LLM assistance, which could be manual (e.g., via a web interface) or through an API-based workflow. 

Validation is confirmatory at this stage. When manually annotated data are available, researchers can reserve a held-out sample for final validation, comparing LLM outputs with human evaluations using standard metrics (e.g., precision, recall, F1). Otherwise, consistency can be assessed through targeted spot checks and inspection of model justifications. 

\section{A Case Study}
\begin{figure}
    \centering
    \includegraphics[width=0.6\linewidth]{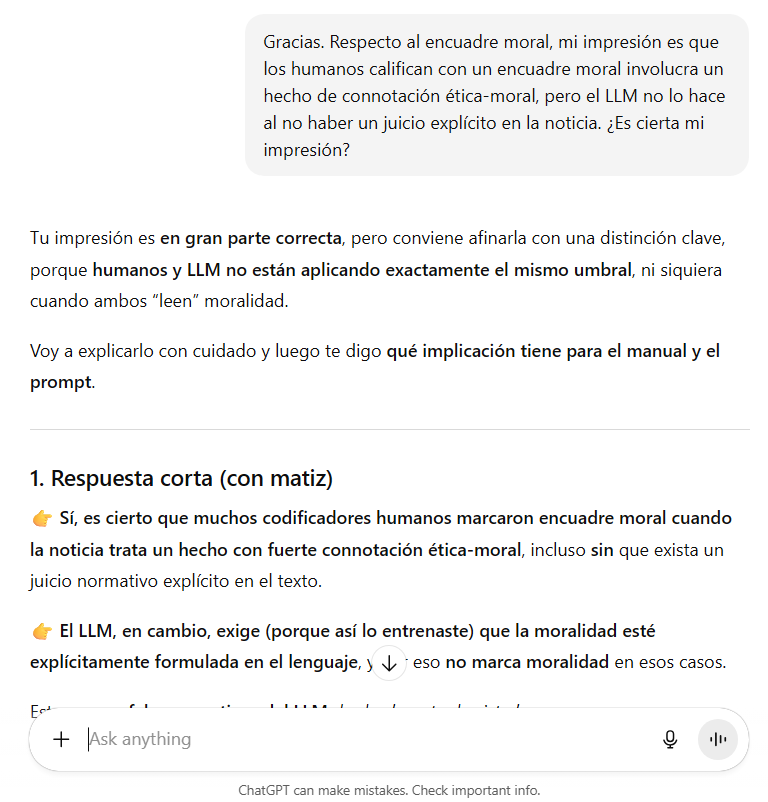}
    \caption{LLM's response to the Initial Prompt.}
    \label{fig:llm_response}
\end{figure}

This section illustrates how the proposed workflow reshaped framing decisions when applied to Valenzuela's \citeyearpar{Valenzuela2023-dy} dataset of Chilean news coverage. The corpus comprises over 3,400 articles from Chilean media outlets, coded according to the four generic frames proposed by Semetko and Valkenburg \citeyearpar{Semetko2000-iz}. We used ChatGPT 5.2 on the browser in December 2025 (See Figure \ref{fig:llm_response}).

\begin{figure}[!htb]
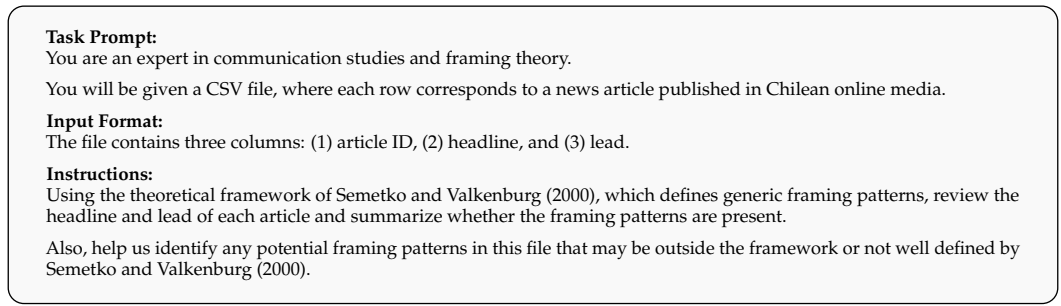

\centering
\begin{tcolorbox}[
    colback=gray!5,
    colframe=black,
    width=\linewidth,
    arc=2mm,
    boxrule=0.5pt,
    breakable
]
\scriptsize
\textbf{Task Prompt:} \\
You are an expert in communication studies and framing theory.

\vspace{4pt}
You will be given a CSV file, where each row corresponds to a news article published in Chilean online media.

\vspace{4pt}
\textbf{Input Format:} \\
The file contains three columns: (1) article ID, (2) headline, and (3) lead.

\vspace{4pt}
\textbf{Instructions:} \\
Using the theoretical framework of Semetko and Valkenburg (2000), which defines generic framing patterns, review the headline and lead of each article and summarize whether the framing patterns are present.

\vspace{4pt}
Also, help us identify any potential framing patterns in this file that may be outside the framework or not well defined by Semetko and Valkenburg (2000).
\end{tcolorbox}
\caption{Initial LLM Codebook Prompt employed in Valenzuela et al.’s study (2017)}
\label{fig:initial_prompt}
\end{figure}

After reviewing the theoretical framework's definitions, we uploaded a spreadsheet with news articles' headlines and leads to ChatGPT, prompting it to summarize framing patterns and identify potential frames outside the original framework. We began the conversation with an initial prompt that included the citation and instructions for the theoretical framework (Figure \ref{fig:initial_prompt}). Without classifying individual articles, the LLM described how well-represented each frame was and identified latent frames that may not align with the original framework. The LLM identified the `conflict' frame as appearing frequently in headlines that mention opposition between actors (e.g., ``Government vs. opposition,'' ``neighbors accuse...'') and use adversarial language. The `responsibility' frame was the next most frequent, showing causal attributions to actors and public institutions resolving problems (e.g., ``The government is responsible...,'' ``... authorities failed...''). The least frequent was `morality,' which the LLM noted \textit{``is rarely explicit; it usually appears diluted within the conflict or responsibility.''} According to the LLM, this frame was related to normative language (e.g., ``outrageous,'' ``unacceptable''), ethical judgments, and discussions on religion and social issues. 

Additionally, the LLM surfaced latent frames characteristic of the corpus: security and order, which overlap with the responsibility frame but emphasize prevention and control rather than blame; exceptional and rare events, which fall outside the human-interest frame and capture surprising, viral, or clickbait-oriented coverage; and risk and emergency events, which were prevalent in Chilean news outlets without a responsible actor. These suggestions and explanations illustrate the workflow's creative potential to both generate new frames and refine existing ones in response to empirical patterns.

For Phase 3, we asked the LLM to identify clear, borderline, and ambiguous cases for each frame. Notable scenarios emerged: for conflict, the LLM flagged sports games (e.g., \textit{``Costa Rica defeated Greece by penalties''}) as ambiguous since a conflict exists but not in the political sense; for human interest, a tsunami alert article was flagged as ambiguous because it conveys human emotions but lacks personalization; for morality, \textit{``Nicolas Maduro blames 'fascism' for violent protests''} was flagged as borderline since moral judgment is present but implicit in political conflict.

Using this information, we refined the initial codebook to address borderline cases. In one iteration, the LLM highlighted that the conflict frame's current version does not cover implicit conflicts (e.g., \textit{``an investigation without an explicit accusation''}) or tensions between realities (e.g., \textit{``scientific truth vs. rumor''}), suggesting the instruction: \textit{``implicit tensions without visible actors should not be coded as conflict.''} For human interest, the LLM noted many articles \textit{``... do not have deep personal stories, but they use a minimal emotional language,''} recommending to edit the original definition with: \textit{``Human interest requires explicit personalization or overt emotionality, not just the presence of an individual.'' }

Importantly, not all distinctions surfaced by the LLM were incorporated into the revised codebook. In some cases, the proposed model introduced differentiations that were empirically observable but theoretically misaligned with the framing research. For instance, the LLM suggested treating rhetorical exaggeration or sensational language as a distinct frame in ambiguous conflict cases. We rejected this proposal because it conflated stylistic devices with framing constructs grounded in problem definition and causal interpretation. Such moments of rejection were analytically productive, as they clarified theoretical boundaries and reinforced the researcher's role in determining which empirical patterns warrant conceptual status. 

Applying the refined codebook to the full corpus also revealed where the original criteria were underspecified. For example, a news article covered the 2014 iCloud Leak scandal involving actress Kate Upton (See Figure \ref{fig:kate_upton_article} for the extract), which discussed privacy violations. Using the original morality definition, this article would likely not be coded as moral, as \textit{``...it lacks explicit references to broader ethical principles, religious values, or normative prescriptions.''} However, contemporary understandings of morality encompass privacy and digital respect. This illustrates how examining LLM justifications reveals definitional ambiguities that require theoretical clarification.

\begin{figure}[!htb]
\centering
\begin{tcolorbox}[
    colback=gray!5,
    colframe=black,
    width=\linewidth,
    arc=2mm,
    boxrule=0.5pt,
    breakable
]
\footnotesize

\textbf{Kate Upton's Lawyer Addresses Controversial Photo Leak}

\vspace{4pt}
\textit{"This is obviously a scandalous violation of the privacy of my client, Kate Upton," her lawyer said.}

\vspace{6pt}
The model was one of the celebrities affected by the massive leak of intimate photos. Model and actress Kate Upton is the latest celebrity to confirm that nude photos of her were stolen and published as part of a massive leak of private images of actresses and singers. "This is obviously a scandalous violation of the privacy of my client, Kate Upton," said Upton's attorney, Lawrence Shire. "We intend to pursue whoever disseminated and/or duplicated these illegally obtained images," he added.

\vspace{4pt}
The leak of the images, which also includes intimate photographs of actresses such as Jennifer Lawrence and Kirsten Dunst, is considered one of the biggest scandals and violations of privacy in history.

\vspace{4pt}
While the individual who leaked the images has not yet been identified, there is speculation that a supposed flaw in Apple's Find My iPhone service could have allowed hackers to access the intimate photos.

\vspace{4pt}
According to specialized sites, a couple of days before the leak, a code was circulating that allowed hackers to access iCloud accounts using a tool that tries all kinds of passwords until the correct one is found, without any warnings.
\end{tcolorbox}
\caption{Ambiguous case in the Morality Frame. Source: Valenzuela et al. (2023), ID: 25801.}
\label{fig:kate_upton_article}
\end{figure}

As the dataset includes manually annotated labels, we also conducted a large-scale evaluation to compare the agreement between the resultant LLM codebook and the coding done originally by human evaluators. In initial iterations, the results indicated a systematic divergence in how moral framing was flagged. Human coders frequently labeled stories involving inherently ethical issues (e.g., abuse, corruption) as moral even without explicit normative language, while the LLM identified moral framing only when moral judgments were explicitly articulated, adhering to Semetko and Valkenburg's \citep{Semetko2000-iz} original definition. A similar divergence emerged with human interest, where human coders required explicit narrative, emotional, or normative cues, whereas the LLM systematically inferred emotional or moral framing from the presence of individuals, accusations, or norm violations. Finally, we observed divergence in the economic consequences frame: the LLM classified stories about sports contracts and corruption under this frame, whereas human coders did not, as monetary figures did not organize the stories' framing. This insight later informed refinements to the definitions and examples of the prompt (See Figure \ref{fig:full_prompt} and \ref{fig:full_prompt_2} for the final LLL codebook's version). 

Table \ref{tab:results} presents the final results of our LLM codebook version. These metrics are reported for transparency rather than optimization. While further adjustments could improve scores, the LLM's justifications offer insights into disagreements that can inform codebook revision done by humans, an opportunity that traditional models do not provide.

\begin{table*}[!htb]
\centering
\scriptsize
\resizebox{\textwidth}{!}{
\begin{threeparttable}
\begin{tabular}{l|cccc|cccc|cccc|cccc}
\toprule
 & \multicolumn{4}{c|}{\textbf{Conflict} (32.74\%)} 
 & \multicolumn{4}{c|}{\textbf{Economic} (16.12\%)} 
 & \multicolumn{4}{c|}{\textbf{Human Interest} (43.88\%)} 
 & \multicolumn{4}{c}{\textbf{Morality} (22.2\%)} \\
\textbf{Model} 
& Acc & Pr & Re & F1 
& Acc & Pr & Re & F1 
& Acc & Pr & Re & F1 
& Acc & Pr & Re & F1 \\
\midrule
Random 
& 0.55 & 0.49 & 0.49 & 0.49 
& 0.73 & 0.51 & 0.51 & 0.51 
& 0.52 & 0.51 & 0.51 & 0.51 
& 0.64 & 0.49 & 0.49 & 0.49 \\

Naive Bayes 
& 0.74 & 0.71 & 0.71 & 0.71 
& 0.83 & 0.65 & 0.58 & 0.60 
& 0.60 & 0.60 & 0.60 & 0.59 
& 0.74 & 0.60 & 0.59 & 0.59 \\

TF-IDF + RF
& 0.79 & 0.80 & \textbf{0.71} & \textbf{0.73} 
& 0.87 & \textbf{0.90} & 0.60 & 0.63 
& \textbf{0.64} & 0.65 & \textbf{0.61} & \textbf{0.60} 
& 0.78 & \textbf{0.74} & 0.51 & 0.47 \\

GPT-5 mini 
& \textbf{0.79} & \textbf{0.81} & 0.70 & 0.71 
& \textbf{0.88} & 0.86 & \textbf{0.66} & \textbf{0.71} 
& 0.63 & \textbf{0.69} & 0.58 & 0.54 
& \textbf{0.79} & 0.70 & 0.56 & 0.56 \\

\bottomrule
\end{tabular}
\begin{tablenotes}
\tiny
\item Note: We report the macro average scores for Precision, Recall, and F1. Based on the ground-truth data, Precision measures how often a model’s prediction of a specific frame was actually correct; Recall assesses how effectively the model identified all articles containing that frame; and the F1 score balances precision and recall into one overall performance measure. The proportions of news articles per frame based on the ground-truth data are in parentheses. In the dictionary-based approach, we utilized TF-IDF to extract features from the news articles and created two models using Random Forest to classify the news articles’ frames.
\end{tablenotes}
\end{threeparttable}}
\caption{Summary of the results using Valenzuela et al.’s database (2017). We report the accuracy (Acc), precision (Pr), recall (Re), and F1 scores for each frame's model.}
\label{tab:results}
\end{table*}

\section{Methodological Recommendations and Limitations}
This approach demonstrates that LLMs can be productively integrated into deductive content analysis as analytic collaborators rather than just automated classifiers. As news production continues to grow in volume, diversity, and pace, journalism researchers face the dual challenge of scaling analysis while maintaining conceptual clarity and theoretical accountability. The proposed workflow addresses this challenge by utilizing structured interaction with LLMs to externalize decision rationales, identify latent frames, surface ambiguities, and facilitate the iterative codebook refinement.

\subsection{Contributions and Advantages}
The primary contribution of this approach is not scalability or efficiency, but the introduction of a creative, theory-facing workflow for deductive framing analysis. By positioning interaction between researchers and LLMs as a site for analytic reflection rather than automated classification, the workflow enables theoretical refinement during analysis rather than treating codebook definitions as settled in advance. In particular, LLM-researcher dialogues can surface hidden assumptions embedded in frame definitions that might otherwise remain implicit. For example, divergence between human coders and the LLM in morality framing revealed that the original codebook implicitly required explicit normative language, whereas human coders applied broader interpretative criteria. Such divergences are not merely reliability problems; they expose conceptual ambiguities in how frames are defined and operationalized over time, creating opportunities for more transparent, adaptable, and theoretically robust framing analysis.

Second, the methodology helps clarify frame boundaries. Borderline cases (e.g., sports conflicts versus political conflicts; money topics in corruption cases or sports negotiations) pose a persistent challenge in framing research, as they necessitate explicit theorization of where one frame ends and another begins. Similarly, the workflow can help researchers identify latent or emergent frames that established frameworks miss, as the LLM can be employed to surface frames or topics that do not necessarily fit with the original terms in the theoretical frameworks. The morality/privacy example suggests that frame definitions may be historically contingent; what counts as ``moral'' in contemporary digital contexts may differ from what was considered moral in the pre-Internet era.  

LLMs also offer several methodological advantages over earlier computational approaches. They allow researchers to embed theoretical definitions, decision rules, and illustrative cases directly into prompts and to revise them iteratively without model training or feature engineering. Importantly, LLMs make reasoning inspectable through natural-language explanations, enabling researchers to interrogate borderline cases in ways that mirror coder training and consensus-building practices. Throughout, researchers retain interpretative authority by defining theoretical boundaries, selecting representative cases, revising criteria, and determining when the codebook is sufficiently stabilized. LLMs function as ``augmentative interlocutors'' that support reflection, comparison, and refinement, rather than as substitutes for human judgment. As multimodal journalism research expands, automated framing coding may also extend to the integrated analysis of text, images, audio, and video.

\subsection{Limitations and Challenges}
While best suited for deductive analysis, this approach is less applicable to inductive coding, which requires careful design and researcher oversight rooted in domain expertise. Future work should explore how LLM-assisted exploration might be combined with human-led inductive analysis in more systematic ways (e.g., open coding conducted by the LLM and axial coding by the researcher).
The method also departs from traditional multi-coder reliability paradigms. Iterative interrogation can approximate aspects of coder consensus, but LLM outputs do not substitute for diverse human interpretations. Using multiple models or prompting strategies may introduce variety, but it remains an imperfect analog of human disagreement and discussions.

Finally, this workflow inherits biases from both researchers and LLMs \citep{Dai2024-ns}. Prompt design and example selection shape model behavior, while LLMs reflect biases embedded in their training data. Researchers risk overfitting, achieving accuracy on initial cases but poor generalization to unseen or new articles. Transparency in documenting codebook revisions, decision rationales, and validation choices is therefore essential, as recognizing the limitations of this approach is relevant to maintaining reflexivity and reproducibility.

\section{Conclusion}
This article introduced a workflow for integrating LLMs into framing coding as analytic collaborators rather than automated classifiers. The proposed workflow treats interaction, disagreement, and revision as productive analytic moments that support methodological creativity in theory-driven research, enabling scholars to surface and refine the theoretical assumptions embedded in their codebooks. The case study demonstrated that LLM-researcher interaction yields more than efficient coding, as the divergences between LLM and human classifications revealed implicit theoretical assumptions and borderline cases that forced explicit theorization of frame boundaries. We encourage journalism and communication researchers to approach LLMs as analytic interlocutors that make theoretical criteria visible and revisable throughout the analysis, thereby expanding the creative possibilities of content analysis beyond efficiency gains.

\section*{Acknowledgments}
This work was partially supported by the Agencia Nacional de Investigación y Desarrollo under Grants ICN1\_002, ICN2021\_004, FB210017, 11140897, CM-ICN17\_002, ICM-NCS2022\_046, and 1231582; and Microsoft Research Accelerating Foundation Models Research grant program. 

\section*{Declaration of Competing Interests}
The authors declare that they have no known competing financial interests or personal relationships that could have influenced the work reported in this paper.

\section*{Use of Generative AI Disclosure}
In preparing and editing this article, we utilized Generative Artificial Intelligence (AI) tools as an aid in refining language and enhancing clarity. We employed OpenAI's ChatGPT 5.2 and ClaudeAI's Opus 4.5 models to revise paragraphs and sentences. We also used LLMs in our methodology and analyses. While we employed this tool in the editing process, we reviewed and approved all content and final decisions, ensuring that the article accurately reflected our research and viewpoints. We acknowledge the role of AI in streamlining the editing process and take full responsibility for the content presented.

\bibliography{colm2026_conference}
\bibliographystyle{colm2026_conference}

\appendix
\section{Appendix}
\begin{figure}[!htb]
\centering
\begin{tcolorbox}[
    colback=gray!5,
    colframe=black,
    width=\linewidth,
    arc=2mm,
    boxrule=0.5pt,
    breakable
]
\begin{lstlisting}
You are an expert in communication studies and framing theory.
Your task is to analyze journalistic news articles published in Chilean online media.
The first line contains the news headline. The second line contains the subheadline (dek).
The rest of the document contains the body of the news article.

You must identify the presence of news frames using adjusted definitions aligned with Semetko & Valkenburg (2000) and with how human coders apply these frames in practice.

==============================
GENERAL INSTRUCTIONS (MANDATORY)
==============================
- Analyze only the headline and the subheadline.
- Use conservative coding: mark a frame only if it is clearly present.
- An article may contain multiple frames.
- Do not infer intentions or evaluations that are not explicit in the text.
- Answer all questions using binary values (1 = yes, 0 = no).

==============================
FRAMES, DEFINITIONS, AND EXAMPLES
==============================

CONFLICT
Definition:
The news emphasizes an explicit disagreement or confrontation between identifiable actors,
presenting opposing positions in the text.

Examples (YES):
- ``Government and opposition clash over tax reform.''
- ``UDI accuses Nueva Mayor\'ia of mismanagement.''

Examples (NO):
- ``Minister announces new measure.''
- ``Authorities investigate the case.''

Questions:
- conflicto_pregunta1: Does the news describe an explicit disagreement or confrontation between actors?
- conflicto_pregunta2: Are at least two opposing positions presented in the text?

------------------------------

ECONOMIC CONSEQUENCES
Definition:
The news is primarily organized around collective or systemic economic effects
(macro, sectoral, or state-level), and the economic impact explains its public relevance.

Examples (YES):
- ``Chilean exports fell 1.2% due to lower copper prices.''
- ``Transit ban will generate million-dollar losses for the transport sector.''
- ``Government announces increase in public spending.''

Examples (NO):
- ``Player signs a million-dollar contract.''
- ``Prosecutor's office investigates $3 billion fraud.''
- ``Club negotiates record-breaking signing.''

Questions:
- economico_pregunta1: Is economic impact the main frame for understanding the news?
- economico_pregunta2: Is the relevance of the event explained mainly by its collective or systemic economic effects?

------------------------------

HUMAN INTEREST
Definition:
The news explicitly develops a personal story or individual experience with a clear emotional load
(empathy, suffering, drama, or private life).

Examples (YES):
- ``Family recounts the tragedy after a fire destroyed their home.''
- ``Couple shares their experience after losing 200 kilos.''

Examples (NO):
- ``Lawmakers meet with the minister.''
- ``Government announces new measures.''

\end{lstlisting}
\end{tcolorbox}
\caption{Final LLM Codebook Prompt employed in Valenzuela et al.'s study (2017)}
\label{fig:full_prompt}
\end{figure}

\begin{figure}[!htb]
\centering
\begin{tcolorbox}[
    colback=gray!5,
    colframe=black,
    width=\linewidth,
    arc=2mm,
    boxrule=0.5pt,
    breakable
]
\begin{lstlisting}
Questions:
- interes_humano_pregunta1: Does the news develop a personal story or individual experience with explicit emotional content?
- interes_humano_pregunta2: Is emotionality or private life central to the narrative, and not merely the mention of people?

------------------------------

MORALITY
Definition:
The news explicitly articulates an ethical or normative judgment, evaluating events or actors
in terms of moral values, ethical principles, or right/wrong.
The ethical connotation of the event alone is not sufficient.

Examples (YES):
- ``Authorities describe the institution's actions as unacceptable.''
- ``Church rejects euthanasia for ethical reasons.''

Examples (NO):
- ``Prosecutor's office investigates abuse.''
- ``Federation risks sanction due to doping.''

Questions:
- moralidad_pregunta1: Does the text use explicit normative or ethical language to judge events or actors?
- moralidad_pregunta2: Are moral values, ethical principles, or notions of right/wrong explicitly invoked?

------------------------------

OUTPUT FORMAT (MANDATORY)
Return ONLY a valid JSON with the following structure:

{
  "nombre_del_archivo.txt": {
    "conflicto_pregunta1": 0,
    "conflicto_pregunta2": 0,
    "conflicto_justificacion_pregunta1": "...",
    "conflicto_justificacion_pregunta2": "...",
    "economico_pregunta1": 0,
    "economico_pregunta2": 0,
    "economico_justificacion_pregunta1": "...",
    "economico_justificacion_pregunta2": "...",
    "interes_humano_pregunta1": 0,
    "interes_humano_pregunta2": 0,
    "interes_humano_justificacion_pregunta1": "...",
    "interes_humano_justificacion_pregunta2": "...",
    "moralidad_pregunta1": 0,
    "moralidad_pregunta2": 0,
    "moralidad_justificacion_pregunta1": "...",
    "moralidad_justificacion_pregunta2": "..."
  }
}

CRITICAL FINAL RULE
- Do not include any text outside the JSON.
- In the "justificacion" fields, provide your reasons for the evaluation of each question.
- Do not add comments or headings.
- Verify that the JSON is valid before submitting it.

File: {filename}

Text:
"""
{text}
"""
\end{lstlisting}
\end{tcolorbox}
\caption{(Continued) Final LLM Codebook Prompt employed in Valenzuela et al.'s study (2017)}
\label{fig:full_prompt_2}
\end{figure}

\end{document}